\documentclass[journal]{IEEEtran}

\IEEEoverridecommandlockouts
\usepackage{amsmath}

\usepackage{amsmath}
\usepackage{booktabs}
\usepackage{cite}
\usepackage{calc}
\usepackage{amsmath,amssymb,amsfonts}
\usepackage{mathrsfs}
\usepackage{graphicx}
\usepackage{textcomp}
\usepackage{xcolor}
\usepackage{mathtools}
\usepackage{amsmath} 
\usepackage{amssymb}
\usepackage{amsfonts}
\usepackage{verbatim} 
\usepackage{bm,color,soul}
\usepackage{algorithm}
\usepackage{algorithm}
\usepackage{algorithmicx}
\usepackage{algpseudocode}
\usepackage{array}
\usepackage[caption=false,font={footnotesize,rm},labelfont=rm,textfont=rm]{subfig}
\usepackage{url}
\usepackage{balance}
\usepackage{hyperref}
\usepackage{mathtools}
\usepackage{amsmath} 
\usepackage{amssymb}
\usepackage{amsfonts}
\usepackage{verbatim} 
\usepackage{bm,color,soul}
\usepackage{cite}
\usepackage{stfloats}
\usepackage{subeqnarray}
\usepackage{cases}
\usepackage{amsthm}
\usepackage{makecell} 
\usepackage{enumerate}
\usepackage{multirow}
\usepackage{enumitem}

\newcommand{\rnum}[1]{\uppercase\expandafter{\romannumeral #1\relax}}
\providecommand{\lemmaname}{Lemma}
\providecommand{\propositionname}{Proposition}
\providecommand{\remarkname}{Remark}

\usepackage{bm}

\ifCLASSINFOpdf
\else
\fi

\begin{document}
\title{Synergizing CKM and ISAC for Environment-Aware Wireless Networks}

\author{Juncong~Zhou,
        Yilong~Chen,
        Yuan~Guo,
        Zixiang~Ren,
        Han~Hu,
        Jie~Xu,~\IEEEmembership{Fellow,~IEEE,}
        and~Rui~Zhang,~\IEEEmembership{Fellow,~IEEE}% <-this % stops a space
\thanks{J. Zhou, Y. Chen, Y. Guo, and J. Xu are with the
 School of Science and Engineering (SSE), the Shenzhen Future Network of Intelligence Institute (FNii-Shenzhen), and the Guangdong
 Provincial Key Laboratory of Future Networks of Intelligence, The
 Chinese University of Hong Kong (Shenzhen), Shenzhen, Guangdong 518172,
 China (e-mail: juncongzhou@link.cuhk.edu.cn; chenyilong@cuhk.edu.cn; 
 guoyuan@cuhk.edu.cn; xujie@cuhk.edu.cn). \\Yuan Guo and Jie Xu are the corresponding authors.}
 \thanks{H. Hu is with the School of Information and Electronics, Beijing Institute
 of Technology, Beijing 100081, China (e-mail: hhu@bit.edu.cn).}
 \thanks{ Z. Ren and R. Zhang are with the Department
 of Electrical and Computer Engineering, National University of Singapore,
 Singapore 117583 (e-mail: zixiangren@nus.edu.sg; elezhang@nus.edu.sg).}}% <-this % stops a space

\maketitle

\begin{abstract}
Channel knowledge map (CKM) and integrated sensing and communications (ISAC) have emerged as two key technologies for sixth-generation (6G) and beyond wireless networks, enabling a new paradigm toward environment-aware system design. CKMs provide static or stochastic wireless environment information \emph{a priori} by learning location-dependent mappings from transceiver positions to channel knowledge of various types, whereas ISAC exploits wireless signals to sense environmental objects in real time. CKM and ISAC have primarily been studied separately in prior works, while their integration offers a promising approach for enhancing the environment awareness of 6G networks and further improving both sensing and communication capabilities. On the one hand, sensing can support CKM construction by providing location-tagged measurements and richer environmental observations, while also enabling real-time CKM refinement in dynamic environments. On the other hand, CKM can enhance sensing by offering prior environmental information for clutter mitigation, and CKM and sensing information can be jointly leveraged to improve communications. This article provides an overview of the complementary interactions between CKMs and ISAC, with particular emphasis on three design approaches: sensing-assisted CKM construction and refinement, CKM-assisted sensing, and joint CKM- and sensing-enhanced communications. We also discuss future research directions for achieving environment-aware wireless networks.
\end{abstract}


\IEEEpeerreviewmaketitle

\section{Introduction}
Sixth-generation (6G) and beyond wireless networks are expected to support emerging applications such as embodied AI agents, autonomous driving, extended and mixed reality, and digital twins. To enable these applications, future networks will rely on ultra-dense base station (BS) deployments, extremely large-scale antenna arrays (ELAAs), and ultra-wideband transmissions to provide enhanced communication, sensing, computing, and AI capabilities\cite{Systematic6G}. As wireless systems become increasingly complex and operate in highly dynamic environments, acquiring accurate channel information is becoming a fundamental challenge for both network planning and operation.

\begin{figure*}[t]
\centering
\includegraphics[width=0.8\linewidth,trim=176bp 16bp 84bp 89bp,clip]{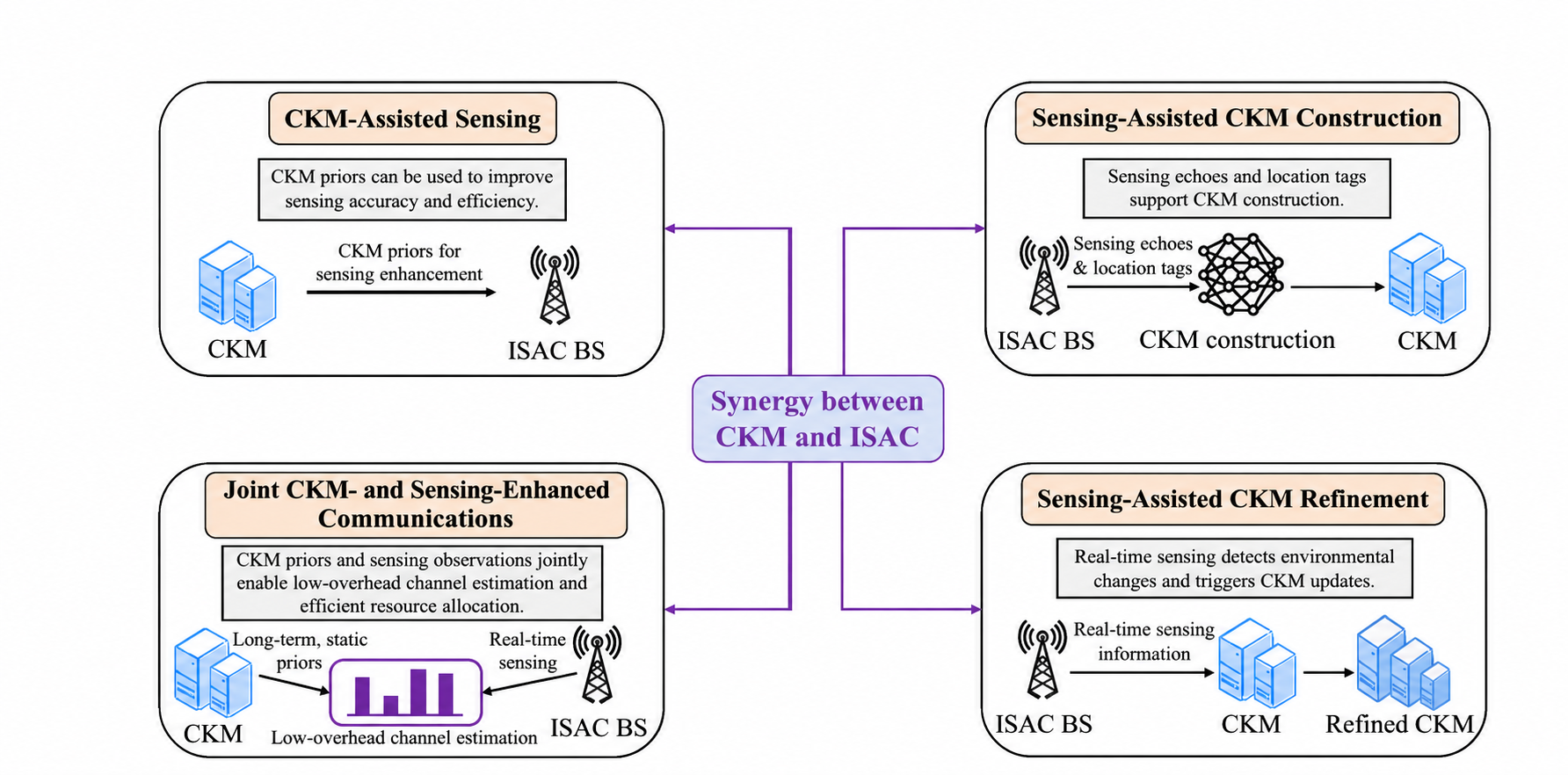}
\caption{Illustration of the synergy between ISAC and CKM, including CKM-assisted sensing, sensing-assisted CKM construction and refinement, and joint CKM- and sensing-enhanced communications.}
\label{clear_illustration}
\end{figure*}

Conventional wireless systems primarily rely on real-time channel training and feedback to acquire channel state information (CSI). However, as channel dimensions continue to increase and network scales expand, such environment-unaware approaches incur substantial signaling overhead and computational complexity. Moreover, they typically ignore rich environmental information and instead rely on simplified statistical channel models or static assumptions about the propagation environment\cite{zeng2021toward}. As a result, there is growing interest in environment-aware wireless networks, where communication systems proactively learn, maintain, and exploit environmental information to facilitate channel knowledge acquisition and network optimization\cite{zeng2024tutorial,CKMandUAVplacement}.

Among the enabling technologies for environment-aware networking, channel knowledge map (CKM)\cite{zeng2024tutorial} and integrated sensing and communications (ISAC)\cite{liu2022integrated} have attracted significant attention. A CKM establishes a location-dependent functional mapping between transceiver positions and channel knowledge, allowing channel information to be predicted and accessed when needed. By providing channel knowledge {\it a priori}, CKMs can reduce, and in some cases even eliminate, the reliance on extensive real-time channel measurements. Meanwhile, ISAC has emerged as a key 6G technology that enables wireless systems to simultaneously communicate and sense the surrounding environment using shared radio resources. Through real-time sensing of objects and propagation environments, ISAC can support environmental reconstruction, wireless imaging, localization, and communication enhancement while improving spectrum utilization efficiency.

Although CKMs and ISAC have largely been investigated as separate research directions in the literature, they are inherently connected through their common dependence on wireless environments and propagation characteristics. Their integration therefore offers a promising opportunity to further enhance the environmental awareness of future wireless networks. On the one hand, ISAC can significantly facilitate CKM construction and refinement. Existing CKMs are typically built offline from large volumes of location-tagged channel measurements. ISAC-based localization can naturally provide the spatial labels required for associating channel measurements with physical locations. Furthermore, sensing observations contain rich environmental information that can complement conventional communication measurements during CKM construction. More importantly, while a CKM may become outdated when the environment changes after deployment, ISAC can help continuously monitor environmental dynamics and provide timely updates for CKM refinement, thereby maintaining map accuracy over time.

On the other hand, CKMs can substantially improve sensing and communication performance. In many sensing applications, repeatedly sensing quasi-static environmental structures introduces unnecessary overhead, while echoes from static objects generate clutter that may obscure dynamic targets and degrade sensing accuracy, especially in non-line-of-sight (NLoS) scenarios. By providing prior knowledge of the surrounding environment and propagation characteristics, CKMs can help suppress clutter, improve target detection, and facilitate environmental interpretation\cite{xu2024channel}. In addition, the combination of long-term channel priors provided by CKMs and real-time observations obtained through sensing can enable more accurate and lower-complexity channel estimation, thereby benefiting adaptive resource allocation and efficient communications.

Motivated by these complementary capabilities, this article provides a systematic overview of environment-aware wireless networks enabled by the synergy between CKMs and ISAC. Specifically, we focus on three representative design approaches: sensing-assisted CKM construction and refinement, CKM-assisted sensing, and joint CKM- and sensing-enhanced communications. {The overall framework and key CKM--ISAC interactions are illustrated in Fig.~\ref{clear_illustration}, with sensing-assisted CKM construction and refinement shown as separate functional blocks.} Note that the ISAC-based localization capability can also facilitate CKM construction via providing devices' location tags. This capability has been widely investigated in the literature \cite{liu2022integrated,isaclocalization} and is therefore not discussed in detail in this article. {Complementing existing CKM tutorials and ISAC surveys, which largely treat the two topics separately, this overview highlights the fundamental connections between CKMs and ISAC, their mutual benefits, and promising research directions for future environment-aware wireless networks. A case study on constructing a 6D CKM from sensing echoes is further presented to illustrate the feasibility of sensing-assisted CKM construction.}

\section{Sensing-Assisted CKM Construction and Refinement}

\begin{table}[t]
\centering
\caption{Channel Power Gain Prediction Performance for 6D CGM Construction Based on Sensing Echoes Across Three Scenarios}
\begin{tabular}{|c|c|c|c|}
\hline
Scenarios & Metric & MLP  & \textbf{BiWGS} \\
\hline
\multirow{1}{*}{Conf. room} 
& MAE & 7.42 dB & \textbf{5.86 dB} \\
\hline
\multirow{1}{*}{Bedroom} 
& MAE & 11.90 dB & \textbf{5.40 dB} \\
\hline
\multirow{1}{*}{Office} 
& MAE & 21.69 dB & \textbf{10.54 dB}\\
\hline
\end{tabular}
\label{6D_isactable}
\end{table}

\begin{figure*}[t]
\centering
\includegraphics[width=0.8\linewidth]{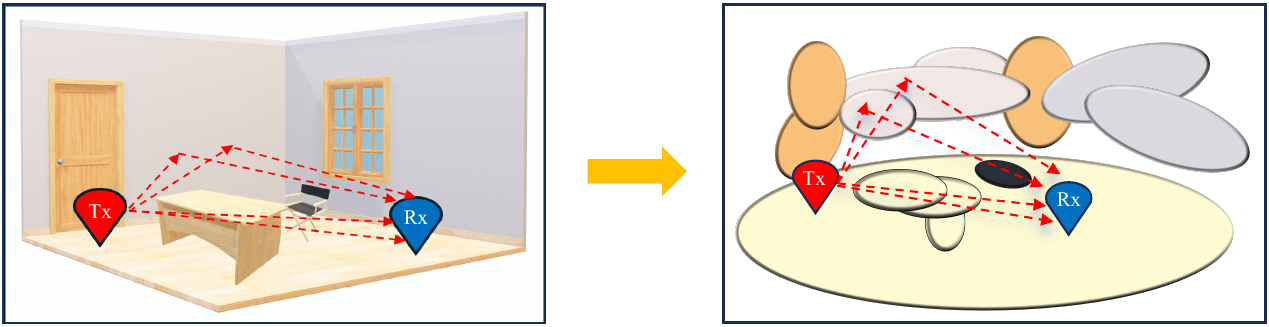}
\caption{Illustration of the bidirectional Gaussian ellipsoid representation in BiWGS. The physical wireless environment with multipath propagation between Tx and Rx is represented as a collection of learnable Gaussian ellipsoids. Each ellipsoid acts as a primitive that captures bidirectional scattering and attenuation properties.}
\label{BiWGS_illustration}
\end{figure*}

\begin{table*}[t!]
\centering
\caption{Comparison of Mean Absolute Error for 6D CGM Construction Using Sensing Echoes and Bistatic Measurements Under Different Training-Set Sizes}
\label{tab:training_size_comparison_double_column}
\renewcommand{\arraystretch}{1.2}
\resizebox{\textwidth}{!}{
\begin{tabular}{|c|cc|cc|cc|}
\hline
\multirow{2}{*}{Training-Set Size} 
& \multicolumn{2}{c|}{Conf. room} 
& \multicolumn{2}{c|}{Bedroom} 
& \multicolumn{2}{c|}{Office} \\
\cline{2-7}
& Sensing Echoes & Bistatic Measurements
& Sensing Echoes & Bistatic Measurements
& Sensing Echoes & Bistatic Measurements \\
\hline
1.25\% & 7.28 dB  & 6.09 dB  & 6.63 dB  & 6.04 dB  & 10.90 dB & 10.85 dB \\
2.5\%  & 6.80 dB  & 5.61 dB  & 5.98 dB  & 5.01 dB  & 10.68 dB & 10.42 dB \\
12.5\% & 6.59 dB  & 4.87 dB  & 5.72 dB  & 4.78 dB  & 10.57 dB & 10.08 dB \\
50\%   & 5.99 dB  & 4.79 dB  & 5.49 dB  & 4.61 dB  & 10.59 dB & 9.90 dB \\
75\%   & 5.97 dB  & 4.86 dB  & 5.47 dB  & 4.57 dB  & 10.56 dB & 9.82 dB \\
100\%  & 5.85 dB  & 4.75 dB  & 5.40 dB  & 4.55 dB  & 10.54 dB & 9.85 dB \\
\hline
\end{tabular}
}
\end{table*}

This section considers the construction and refinement of CKMs assisted by ISAC. Depending on the dimensionality of the transceiver-location inputs, CKMs can be broadly categorized into BS-to-any (B2X) and any-to-any (X2X) types \cite{zeng2024tutorial}. A B2X CKM maps the location of a mobile user to the corresponding channel knowledge with respect to a fixed-position base station (BS), whereas an X2X CKM characterizes channel knowledge for arbitrary transmitter (Tx)--receiver (Rx) pairs by taking the locations of both nodes as inputs. Depending on the type of channel knowledge being represented, a CKM may take the form of a channel gain map (CGM), channel state information (CSI) map, channel angle map (CAM), or other variants. Furthermore, CKMs can be implemented using different representations, including spatial databases, image-based representations, and deep-learning models\cite{renWirelessCom}.

\subsection{Sensing-Assisted CKM Construction}

A fundamental challenge in CKM construction is the acquisition of sufficient training data. Existing CKMs are typically built from large volumes of location-tagged communication measurements, which often require coordinated Tx and Rx devices as well as accurate position labeling. This challenge becomes particularly pronounced for X2X CKMs, where channel measurements must be collected across numerous Tx--Rx pairs, resulting in significant deployment complexity and data-acquisition overhead.

ISAC provides a promising opportunity to address this challenge in X2X CKM construction\footnote{{The design is also applicable to B2X CKM construction for a fixed BS as  a specific case of X2X CKM, in which only monostatic channel measurements at multiple ISAC device locations are available.}}. {In particular, we consider the monostatic ISAC setup, in which Tx and Rx are co-located at each sensing location. In this case, although moving the sensing node does not provide independently varying Tx--Rx pairs, it is worth noting that the sensing and communication channels are governed by the same physical environment and often involve common propagation paths and scatterers.} Therefore, the sensing echoes inherently contain information about the underlying propagation characteristics for bistatic communication channels  \cite{liu2023shared}. As a result, sensing measurements can serve as an alternative or complementary source of training data for X2X CKM construction. 

Compared with conventional bistatic communication measurements, sensing-assisted CKM construction offers several advantages. First, monostatic sensing significantly reduces the reliance on coordinated Tx--Rx measurement infrastructure, thereby simplifying data collection. Second, the localization burden can be greatly reduced, as only the sensing-capable transceiver needs to be tracked during measurement acquisition. Collectively, these advantages make sensing-assisted CKM construction an attractive infrastructure-light solution for future large-scale wireless networks.

{To validate this idea, we consider an illustrative case study based on the bidirectional wireless Gaussian splatting (BiWGS) framework proposed in \cite{zhou20256d}.} BiWGS is a 6D X2X CKM representation that models the wireless environment using a collection of bidirectional Gaussian primitives, whose parameters are learned from measurement data and subsequently used to infer channel knowledge at previously unseen Tx--Rx locations. The Gaussian ellipsoid representation of BiWGS is illustrated in Fig.~\ref{BiWGS_illustration}. Conventionally, BiWGS is trained using bistatic communication measurements collected from Tx--Rx pairs. {Nevertheless, we notice that the shared propagation characteristics of sensing and communication links provide a physical basis for predicting bistatic channels from monostatic echoes. Therefore, by learning the environmental scattering and attenuation characteristics embedded in monostatic sensing echoes, BiWGS can still infer bistatic channel power gains at different Tx--Rx locations.} We use monostatic sensing echoes to train BiWGS and construct an X2X CKM for bistatic communication links.

In particular, in this case study, we consider the construction of a 6D CGM from monostatic sensing echoes in three simulated 3D wireless environments{, namely a conference room, a bedroom, and an office}, as summarized in Table~\ref{6D_isactable}. The datasets are generated using the Sionna ray-tracing simulator\footnote{Sionna is an open-source GPU-accelerated library for end-to-end wireless system simulation, available at \url{https://developer.nvidia.com/sionna}.}. {For the comparison in Table~\ref{6D_isactable}, both BiWGS and the multilayer perceptron (MLP) benchmark, which represents a neural-network-based CKM, are trained on monostatic sensing measurements collected with co-located Tx and Rx. Both models are tested on bistatic communication channel power gains from the same environments. The training and test sets each contain 5,349, 7,337, and 6,000 samples for the conference room, bedroom, and office, respectively.} The mean absolute error (MAE) measures the average absolute difference between the predicted and ground-truth channel power gains on the dB scale, where a lower value indicates more accurate prediction. As shown in Table~\ref{6D_isactable}, BiWGS consistently outperforms the MLP-based benchmark, which verifies the feasibility of constructing a 6D CGM from sensing echoes. Nevertheless, the MAE values of BiWGS remain non-negligible because monostatic sensing echoes provide indirect information for bistatic channel power gain prediction, suggesting that further improving sensing-echo-based CKM construction accuracy remains an important open problem.

\begin{figure*}[t]
\centering
\includegraphics[width=0.7\linewidth]{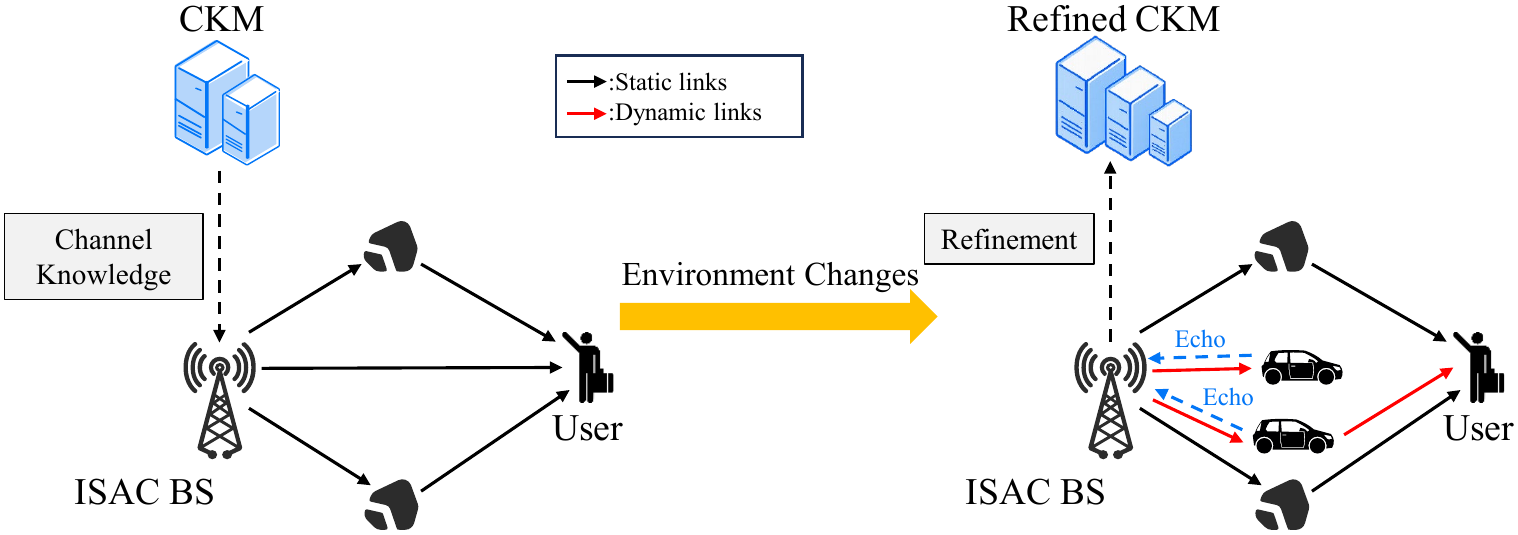}
\caption{Illustration of real-time CKM refinement under a dynamic environment. The black links denote quasi-static communication links that are already captured by the preconstructed CKM. After environment changes, newly appearing or moving objects introduce dynamic communication links, shown by red links, which are not contained in the original CKM. The blue dashed arrows denote real-time sensing echoes from these dynamic objects. By extracting the sensed dynamic information, the CKM can be refined.} \label{fig:CKM_refinement_dynamic_environment}
\label{fig:ISAC_update_CKM}
\end{figure*}

Table~\ref{tab:training_size_comparison_double_column} compares the MAE {of BiWGS trained on either sensing echoes (our proposed sensing-assisted CKM construction) or bistatic communication measurements (conventional design)} under different training-set sizes. {For sensing-echo training, each dataset is split 4:1 into training and sensing-test subsets, with integer rounding where needed. For the conventional design, the  bistatic test set has the same size as the sensing-test subset. Percentages for sensing-echo training refer to the full training subset, while both test sets remain fixed.} Several important observations can be made. First, sensing-only training data enable BiWGS to infer bistatic channel power gains, thereby providing an initial feasibility validation of sensing-assisted X2X CKM construction, although the relatively large MAE values indicate that further accuracy improvement is still needed. Second, under both training settings, the MAE generally decreases as the training-set size increases, although the improvement becomes limited at larger training-set sizes. Third, as expected, CKMs trained using bistatic communication measurements generally achieve lower prediction errors and require fewer training samples. {The performance gap is due to the indirect information provided by sensing echoes for bistatic channel prediction. Bistatic measurements directly characterize the target communication links and therefore provide more informative supervision.} Despite this performance gap, sensing-assisted CKM construction remains attractive in practice, as it eliminates the need for coordinated Tx--Rx measurements and reduces location-labeling requirements, significantly lowering deployment cost and complexity.

This case study highlights a promising paradigm for ISAC-assisted X2X CKM construction, where sensing infrastructure serves not only for environmental perception but also as a scalable source of training data for wireless-environment learning. Future research may investigate how sensing observations can be integrated into other CKM representations and jointly exploited with communication measurements to enable more accurate and scalable CKMs.

\subsection{Sensing-Assisted CKM Refinement in Dynamic Environments}

After a CKM is constructed from historical location-tagged measurements, its accuracy may gradually degrade as the physical environment evolves. For example, the appearance of new blockers, object movements, or changes in reflective surfaces can alter the underlying multipath propagation structure. Such environmental dynamics are difficult to capture using conventional CKM construction approaches, which are typically based on static datasets. Therefore, maintaining CKM accuracy in dynamic environments requires efficient refinement mechanisms that can promptly incorporate newly observed environmental changes with low overhead.

ISAC provides a natural solution to this challenge. By continuously sensing the surrounding environment and detecting dynamic objects, ISAC can identify environmental changes and guide CKM updates in real time. Rather than reconstructing the entire CKM from scratch, refinement only updates the map components affected by the detected changes, thereby reducing computational complexity and enabling efficient adaptation. The basic idea of CKM refinement using real-time sensing is illustrated in Fig.~\ref{fig:ISAC_update_CKM}.

\begin{figure*}[t]
\centering
\includegraphics[width=0.7\linewidth]{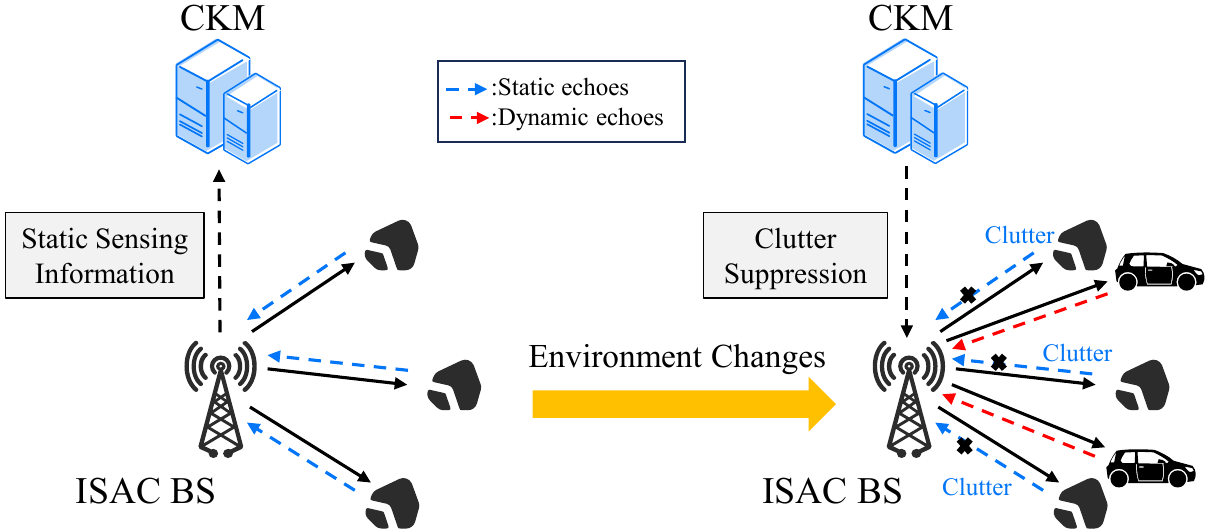}
\caption{Illustration of CKM-assisted sensing enhancement in ISAC. The blue dashed links denote CKM-known static clutter echoes associated with static environmental scatterers, which can be identified and removed through CKM-based clutter suppression. The red dashed links denote dynamic target echoes caused by newly appearing or moving objects, which are not contained in the constructed CKM and need to be obtained through real-time sensing.}
\label{fig:CKM_helps_ISAC}
\end{figure*}

As an example, consider the BiWGS framework, where the wireless environment is represented by a set of 3D Gaussian ellipsoids. In this framework, sensing-assisted refinement is achieved by first detecting environmental changes and characterizing the associated dynamic objects through sensing, and then updating only the ellipsoids affected by these changes while keeping the remaining components unchanged. This localized update mechanism can be viewed as incremental adaptation rather than global retraining, thereby significantly reducing the computational overhead compared with reconstructing the entire CKM. Furthermore, because BiWGS explicitly captures environment-dependent propagation characteristics, the refined model can continuously incorporate newly observed propagation behaviors, improving long-term CKM accuracy and consistency.

Overall, incorporating ISAC into CKM refinement enables an efficient and adaptive framework for maintaining high-fidelity CKMs in dynamic environments, allowing environment-aware wireless networks to continuously evolve with changing propagation conditions.

\section{CKM-enabled Sensing Enhancement}
\label{sec:ckm_helps_sensing}

This section explores how CKMs can enhance sensing performance. By storing environment-dependent channel knowledge, a CKM can provide valuable a priori information about propagation conditions, including blockage locations, clutter characteristics, and multipath propagation structures. An illustrative example of CKM-based clutter suppression for sensing is shown in Fig.~\ref{fig:CKM_helps_ISAC}. By leveraging such prior knowledge, ISAC systems can improve sensing accuracy and efficiency, particularly in challenging NLoS scenarios.

For example, \cite{wu2025you} proposed a CKM-assisted framework for NLoS sensing by reusing a CKM originally constructed for communication purposes. The key idea is to model a passive sensing target as a virtual communication user located at the target position, such that the channel priors stored in the CKM can be transformed into sensing priors and incorporated into a location-dependent likelihood function for target parameter estimation. By exploiting these priors, the proposed approach significantly improves sensing accuracy compared with conventional methods that do not utilize CKM information, especially when line-of-sight (LoS)-based sensing becomes ineffective. Moreover, the results show that richer channel knowledge stored in the CKM leads to further sensing-performance gains.

Another representative example is the clutter angle map (CLAM)-based framework proposed in \cite{xu2024channel}. In this approach, a CLAM stores the dominant clutter angles of arrival associated with different locations. During online sensing, the BS queries the CLAM using the transceiver location and constructs an orthogonal projection to suppress the clutter subspace before applying parameter-estimation algorithms such as multiple signal classification (MUSIC). By shifting part of the clutter-mitigation task from online signal processing to offline channel-knowledge acquisition, this approach improves angle-estimation accuracy while maintaining a complexity level comparable to that of conventional MUSIC.

These examples illustrate the fundamental benefit of CKM-assisted sensing: transforming environmental information acquired offline into useful sensing priors that can reduce uncertainty during online sensing. {Specifically, the NLoS sensing framework incorporates communication channel priors into target parameter estimation. The CLAM-based framework instead exploits stored clutter angles to suppress clutter before parameter estimation.}

Despite these encouraging results, CKM-assisted sensing remains at an early stage and several important challenges remain open. First, most existing studies rely on relatively low-dimensional CKMs. Extending these approaches to more general 6D X2X CKMs, which jointly characterize Tx and Rx states, remains largely unexplored. Such extensions are particularly important for mobile sensing scenarios where both ends of the sensing link may vary over time. Second, existing methods typically assume that the sensing-node or target location used to query the CKM is accurately known. In practice, localization errors may lead to mismatched priors, thereby degrading sensing performance or even introducing misleading clutter-suppression decisions. This calls for uncertainty-aware CKM querying and robust sensing algorithms that explicitly account for location uncertainty. Third, current CKM-assisted sensing frameworks often treat the CKM as a static prior database. However, clutter distributions and propagation characteristics may evolve over time due to environmental changes and moving objects. Future research should therefore investigate how real-time sensing feedback can be leveraged to continuously update CKMs and maintain reliable sensing priors in dynamic environments.

\section{Joint CKM- and Sensing-Enhanced Communications}

This section explores how the synergy between CKMs and sensing can further enhance communication performance in ISAC. Recall that a CKM captures long-term environment and channel information from historical measurements, while sensing provides real-time observations of dynamic environmental changes. By jointly leveraging these complementary sources of information, wireless systems can achieve more accurate and efficient channel estimation, which is particularly important for modern communication systems with large antenna arrays and high-dimensional channels operating under stringent latency and pilot-overhead constraints. {CKM refinement updates the stored map based on sensed environmental changes. Joint communication enhancement instead combines the existing or refined map with current sensing observations to guide online communication decisions. Beyond channel estimation, these decisions include beam management, scheduling, power control, mobility management, and handover.}

Unlike conventional pilot-based approaches that treat wireless channels as unstructured and unknown quantities, CKM- and sensing-enhanced communications exploit both environmental priors and real-time sensing observations to reduce channel uncertainty and facilitate channel reconstruction with significantly lower online overhead. In this sense, CKMs provide long-term environmental memory, while sensing continuously updates the network's knowledge of dynamic propagation conditions.

Building on this idea, \cite{staticanddynamiczengyong} proposed an environment-aware channel estimation framework that combines a CAM with real-time sensing. In this framework, the CAM provides location-dependent angular priors for quasi-static propagation paths, while sensing identifies dynamic objects and updates the candidate angles associated with time-varying paths. Channel estimation is then performed over a significantly reduced set of candidate paths, thereby lowering online processing complexity. However, the CAM mainly serves as a static prior database, and dynamic environmental information is only incorporated during online inference. Consequently, recurring environmental changes may still need to be processed repeatedly rather than being persistently learned and absorbed into the CKM.

A more recent study \cite{guoyuanwork} extended this concept to near-field channel estimation in ELAA-enabled ISAC systems through the introduction of a virtual object map (VOM). The VOM stores BS-visible virtual interaction points associated with static objects, thereby providing an object-level representation of quasi-static multipath propagation. Real-time monostatic sensing is then employed to suppress static clutter and extract a low-dimensional dynamic subspace, which is combined with the VOM-defined static basis for channel reconstruction. The resulting framework improves both channel-estimation accuracy and achievable downlink rate, particularly in short-pilot regimes. Nevertheless, the VOM remains a preconstructed and largely static representation, leaving adaptive map construction and online refinement in evolving environments as important open problems.

These examples illustrate the key advantage of jointly leveraging CKMs and sensing for communications: CKMs provide structured prior knowledge of the propagation environment, while sensing supplies real-time environmental updates. Their integration enables more accurate channel reconstruction with reduced pilot overhead and computational complexity. {Specifically, the CAM-based framework combines stored angular priors with sensing-based angle updates to narrow the candidate path set. The VOM-based framework instead combines an object-level static basis with a sensing-derived dynamic subspace for near-field channel reconstruction.}

Despite these promising advances, several important challenges remain. First, existing solutions are often tailored to specific CKM representations, such as angle-domain maps or virtual-object libraries. Developing a unified framework capable of integrating sensing information with diverse CKM representations remains largely unexplored. Second, most current approaches treat CKMs as static prior databases rather than adaptive environment models that can be continuously updated through sensing observations or communication feedback. Future research should therefore investigate closed-loop designs in which channel estimation, sensing, and CKM refinement are jointly optimized. Third, practical impairments such as sensing errors, localization uncertainty, and processing latency may lead to unreliable priors or delayed environmental updates, potentially degrading communication performance. This calls for uncertainty-aware channel estimation methods that can explicitly evaluate and exploit the reliability of both CKM priors and sensing observations. Finally, future systems should move beyond loosely coupled architectures and develop unified frameworks for pilot transmission, sensing-data processing, CKM querying, CKM updating, and channel reconstruction. Such an integrated design may provide a foundation for robust, low-overhead, and adaptive environment-aware communications in future 6G networks.

\section{Future Directions}
The preceding sections have highlighted several key design approaches that exploit the synergy between CKMs and ISAC for environment-aware wireless networks. These studies suggest that CKMs and ISAC should not be viewed as isolated technologies, but rather as complementary components of a unified environment-aware wireless framework. Despite recent progress, several important research challenges remain open.

\textbf{Large-Scale Multi-BS CKM Construction:}
Although existing studies have demonstrated the feasibility of CKM construction, most validations have been conducted in indoor environments or under single-BS settings. For practical deployment in large-scale outdoor networks, particularly city-scale 6G systems, it is necessary to move beyond the single-BS paradigm and investigate multi-BS CKM construction and management. Such large-scale deployments introduce several new challenges, including CKM partitioning across multiple BSs, boundary management between neighboring CKMs, coordination in overlapping coverage areas, and interference-aware map construction. Moreover, communication and sensing measurements collected by different BSs may need to be jointly exploited to construct consistent and scalable CKMs. Future research should therefore focus on computationally efficient CKM architectures and distributed learning mechanisms for large-scale environment-aware wireless networks.

\textbf{Multimodal Sensing for CKM Construction and Refinement:}
While this article primarily focuses on radio-frequency (RF) sensing, future CKMs will likely benefit from multimodal sensing sources such as radar, cameras, LiDAR, and other environmental sensors. These modalities can provide complementary observations of the physical environment, improving localization accuracy during CKM construction and enabling more reliable detection of environmental changes for CKM refinement\cite{cheng2023intelligent}. {A potential direction is to exploit multimodal large language models (mLLMs)\cite{chen2024mllm} that could fuse RF/radar, camera/LiDAR, pose, historical CKM, and confidence information into object-level representations and candidate CKM updates. Their semantic reasoning capability could help relate the information about objects and environmental changes to the propagation information stored in CKMs.} However, integrating heterogeneous sensing modalities into a unified CKM framework remains challenging due to differences in sensing range, spatial resolution, update frequency, and reliability. Furthermore, some sensing modalities may become unavailable under adverse conditions such as occlusion, poor lighting, or hardware limitations. Future work should investigate robust multimodal fusion architectures that can flexibly exploit available sensing sources while maintaining reliable performance under partial observations. Lightweight implementations that satisfy practical latency, energy, and computational constraints are also needed.

\textbf{Uncertainty-Aware CKM Modeling and Decision Making:}
Since sensing observations and localization outputs are practically imperfect, CKMs constructed from such measurements inevitably contain uncertainty. Ignoring this uncertainty may lead to inaccurate channel predictions and suboptimal communication or sensing decisions. Future research should develop uncertainty-aware CKM representations that characterize both channel knowledge and confidence levels. Such probabilistic CKMs can support robust querying, decision making, and resource allocation under imperfect environmental knowledge. Potential applications include BS deployment, beam management, mobility support, trajectory planning, interference coordination, and CKM-assisted sensing. More broadly, uncertainty-aware CKMs may provide a principled foundation for trustworthy environment-aware wireless networks.

\textbf{Towards Self-Evolving CKM--ISAC Systems:} Another particularly promising long-term direction is the development of self-evolving CKM--ISAC systems. Rather than treating CKMs as static databases and sensing as an independent online process, future networks may continuously integrate sensing observations, communication feedback, and environmental changes to jointly update and refine their environmental knowledge. Such closed-loop learning frameworks could enable CKMs and ISAC to co-evolve over time, forming a continuously improving digital representation of the wireless environment. This capability may become a key enabler of autonomous and intelligent 6G networks.

\section{Conclusion}
This article has presented a systematic overview of the synergy between CKMs and ISAC for achieving environment-aware wireless networks. Rather than treating CKMs and ISAC as separate technologies, we have shown that they can complement and reinforce each other in multiple ways. On the one hand, ISAC can facilitate CKM construction by providing location-tagged measurements and further enable real-time CKM refinement in dynamic environments. On the other hand, CKMs can enhance sensing by providing environment-aware priors for clutter suppression, target detection, and NLoS sensing, while the joint use of CKM priors and sensing observations can further improve channel estimation and communication performance. As future 6G and beyond wireless systems continue to evolve toward larger scales, richer environments, and more stringent performance requirements, the synergy between CKMs and ISAC is expected to play an increasingly important role in enabling intelligent and autonomous wireless networks.

\ifCLASSOPTIONcaptionsoff
  \newpage
\fi

{\scriptsize
\bibliographystyle{ieeetr}
\bibliography{IEEEabrv,ref}
}

\end{document}